\documentstyle[amssymb,11pt]{article}
\renewcommand{\baselinestretch}{1.3}

\input epsf

\begin{document}
\newcount\nummer \nummer=0
\def\f#1{\global\advance\nummer by 1 \eqno{(\number\nummer)}
      \global\edef#1{(\number\nummer)}}


\newcommand{\plabel}{\label}                       
\newcommand{\pcite}{\cite}                         
\newcommand{\pbibitem}{\bibitem}                   

\newcommand{\N}{\mbox{{\rm I \hspace{-0.865em} N}}} 
\newcommand{\Z}{\mbox{$\Bbb Z$}}                    
\newcommand{\R}{\mbox{{\rm I \hspace{-0.86em} R}}}  
                         
\def\Di{\displaystyle}
\def\nn{\nonumber \\}
\def\re{(\ref }
\def\rz#1 {(\ref{#1}) }   \def\ry#1 {(\ref{#1})}
\def\rp#1 {(\ref{#1}) }
\def\i{{\rm i}}
\let\a=\alpha \let\b=\beta \let\g=\gamma \let\d=\delta
\let\e=\varepsilon \let\ep=\epsilon \let\z=\zeta \let\h=\eta \let\th=\theta
\let\dh=\vartheta \let\k=\kappa \let\l=\lambda \let\m=\mu
\let\n=\nu \let\x=\xi \let\p=\pi \let\r=\rho \let\s=\sigma
\let\t=\tau \let\o=\omega \let\c=\chi \let\ps=\psi
\let\ph=\varphi \let\Ph=\phi \let\PH=\Phi \let\Ps=\Psi
\let\O=\Omega \let\S=\Sigma \let\P=\Pi \let\Th=\Theta
\let\L=\Lambda \let\G=\Gamma \let\D=\Delta      

\def\wt{\widetilde}
\def\w{\wedge}
\def\0{\over } \def\1{\vec } \def\2{{1\over2}} \def\4{{1\over4}}
\def\5{\bar } \def\6{\partial }
\def\7#1{{#1}\llap{/}}
\def\8#1{{\textstyle{#1}}} \def\9#1{{\bf {#1}}}

\def\({\left(} \def\){\right)} \def\<{\langle } \def\>{\rangle }
\def\lb{\left\{} \def\rb{\right\}}
\let\lra=\leftrightarrow \let\LRA=\Leftrightarrow
\let\Ra=\Rightarrow \let\ra=\rightarrow
\def\ul{\underline}
                          
\let\ap=\approx \let\eq=\equiv
\let\ti=\tilde \let\bl=\biggl \let\br=\biggr
\let\bi=\choose \let\at=\atop \let\mat=\pmatrix
\def\CL{{\cal L}}\def\CX{{\cal X}}\def\CA{{\cal A}}
\def\CF{{\cal F}} \def\CD{{\cal D}} \def\rd{{\rm d}}
\def\rD{{\rm D}} \def\CH{{\cal H}} \def\CT{{\cal T}} \def\CM{{\cal M}}
\def\CI{{\cal I}} \newcommand{\dR}{\R}
\def\CP{{\cal P}}\def\CS{{\cal S}}\def\C{{\cal C}}
\let\mytilde=\widetilde 


\begin{titlepage}
\renewcommand{\thefootnote}{\fnsymbol{footnote}}
\renewcommand{\baselinestretch}{1.3}
\hfill  TUW - 97 - 13\\
\medskip
\hfill  PITHA - 96/24\\
\medskip

\begin{center}
{\LARGE {A Global View of Kinks in 1+1 Gravity}  
\\ }
\medskip
\vfill
            
\renewcommand{\baselinestretch}{1}
{\large {THOMAS
KL\"OSCH\footnote{e-mail: kloesch@tph.tuwien.ac.at} \\
\medskip
Institut f\"ur Theoretische Physik \\
Technische Universit\"at Wien\\
Wiedner Hauptstr. 8--10, A-1040 Vienna\\
Austria\\
\medskip
\medskip THOMAS STROBL\footnote{e-mail:
tstrobl@physik.rwth-aachen.de} \\ \medskip
Institut f\"ur Theoretische Physik \\
RWTH-Aachen\\
Sommerfeldstr. 26--28, D52056 Aachen\\
Germany\\}}
\end{center}

\setcounter{footnote}{0}
\renewcommand{\baselinestretch}{1}                          
            
\begin{abstract}
  Following Finkelstein and Misner, kinks are non-trivial
  field configurations of a field theory, and different kink-numbers
  correspond to different disconnected components of the space of
  allowed field configurations for a given topology of the base
  manifold. In a theory of gravity, non-vanishing kink-numbers are
  associated to a twisted causal structure. In two dimensions this
  means, more specifically, that the light-cone tilts around
  (non-trivially) when going along a non-contractible
  non-selfintersecting loop on spacetime.
  One purpose of this paper is to construct the maximal extensions of kink
  spacetimes using Penrose diagrams. This will yield surprising insights into
  their geometry but also allow us to give generalizations of some well-known
  examples like the bare kink and the Misner torus.
  However, even for an arbitrary 2D metric with a Killing field we can
  construct continuous one-parameter families of inequivalent kinks. This
  result has already interesting implications in the  flat or deSitter case,
  but it applies e.g.\ also to generalized dilaton gravity solutions.
  Finally, several coordinate systems for these newly obtained kinks
  are discussed.

\medskip

\noindent PACS numbers: 04.20.Gz 02.40.Ky 04.60.Kz

\end{abstract}

\vfill
\hfill July 1997  \\
\end{titlepage}

\renewcommand{\baselinestretch}{1}
\small\normalsize

\section{Introduction}
\plabel{Introduction}

About 40 years ago, Finkelstein and Misner considered 
integer-valued quantities that are conserved during time evolution 
as they are protected by a topological index \pcite{FinMis}. These
quantities, which may be used to characterize a field configuration of
an appropriate field theory, were named ``kinks'' thereafter
\pcite{Fin}. The idea of kinks is simple and by now standard: Suppose
you are dealing with a field theory where the fields take values in a
space $\Omega$ of non-trivial topology (such as, e.g., in a
$\sigma$-model; in gravity this non-triviality results from the
required signature of the metric). Now consider the map $\Phi$ from a
$t=const$ hypersurface $\S$ into $\Omega$ given by the initial values of
the field(s). If $\S$ has non-trivial topology (possibly due to
boundary conditions imposed on the fields on an originally trivial
space), it may well happen that there is more than one homotopy class in
$H(\S,\Omega)$. In this case the initial data, resp.\ $\Phi$, single out
some element $h \in H(\S,\Omega)$. As time evolution is a smooth
deformation of the map $\Phi$, it will not move $\Phi$ out of its original
homotopy class $h$. Thus, $h$ is a conserved quantity and for $h\neq 0$
(0 denoting the trivial homotopy class defined by the constant map)
the field configuration is said to have a kink (characterized by $h$).

In 3+1 gravity on a spacetime $\Sigma \times \dR$ with $\S$ being a
3-sphere, one has $H(\S,\Omega)=\pi_3(\Omega)=\Z$, the group of all integers
\pcite{FinMis,Steen}. The situation is unchanged, if $\S =\dR^3$ and
one requires spacetime to be (appropriately) Minkowskian
asymptotically.  Such spacetimes are characterized by a kink-number
$k\in \Z$ therefore. In subsequent works then it has been shown that
all spacetimes with $k \neq 0$ have a ``twisting light-cone
structure'' and gravitational kinks were in part viewed as ``black
holes without curvature singularities''
(cf., e.g., \pcite{FinPR,McCollum,DeSitterkinks}).
This connection of homotopical considerations with those concerning the 
causal structure becomes most transparent for 1+1 dimensional
spacetimes, which, as often, may serve as a suitable laboratory to
improve one's understanding of the role of kinks in gravitational theories
\pcite{Dunn}. Here $\det g \equiv g_{00}g_{11} - g_{01}^2 \neq 0$
separates $\dR^3$ (the space of real symmetric matrices)
into three regions characterized by the
signatures $(++)$, $(--)$, and $(+-)$, respectively. The latter of
these regions is $\O$. With $\S = S^1$ one obtains 
$H(\S,\Omega)=\pi_1(\Omega) = \Z$, so that there again is a winding
number $k$ characterizing kinks (cf.\ Fig.\ 1 of \pcite{Vasilic} for a
nice illustration).
On the other hand, given an explicit kink metric, such as e.g.\ the ``bare
kink'' \re{barekink}) \pcite{Dunn,Bare}
\[
  g=-\cos2x\,{\rm d}t^2-2\sin2x\,{\rm d}t\,{\rm d}x+\cos2x\,{\rm d}x^2\,\, ,
\]
it is easily verified that the lightcone turns upside down $k$ times when
going from $x=0$ to $x=k\pi$ along a ($t=const$)-line $\S$
(cf.\ Fig.\ 1({\em a\/})). Each such half-turn of the lightcone
clearly defines a non-contractible loop in $\O$,
which may serve as generator of $\pi_1(\Omega)$.
\footnote{Here we always considered spacetimes $\CM$ of the form
  $\CM = \Sigma \times \dR$ with $\Sigma = S^3,\dR^3,S^1,$ or $\dR^1$, all of
  which are parallelizable. In the case of a more general, not parallelizable
  spacetime manifold $\CM$ the metric is a section of a non-trivial bundle
  and the above homotopical considerations have to be modified accordingly.}

In the literature 1+1 kink metrics have often been written down in explicit
coordinates only \pcite{DeSitterkinks,Dunn,Vasilic}.
Their kink nature is then usually
shown by studying the behaviour of the lightcone as sketched briefly
in the example above. Consequently, one rarely finds any global analysis of
the resulting spacetimes 
(the papers \pcite{HawkGiul} being a
positive exception, where the kink-number of the
boundary components of a 3+1 spacetime $\CM$ is shown to be directly
related to the Euler characteristic of $\CM$).
In the present paper we want to fill this gap at least partially.
Moreover, our analysis will lead to a systematic
way of constructing new kink spacetimes. In particular it will enable us to
construct, e.g., a one-parameter family of distinct kinks of given
kink-number for {\em any\/} given 1+1 metric with a local Killing
symmetry. Kinks with $\S = \R$ instead of $\S = S^1$ may also be obtained,
provided only that the metric allows for appropriate asymptotic regions
(say, asymptotically flat or deSitter); however, in these cases the kink-number
will turn out not to be an intrinsic property of the spacetime
but rather a feature of the chosen coordinates.

While there is no problem in just writing down kink-metrics (cf.\ e.g.\
\re{barekink})), the more interesting cases are certainly those where the
metric fulfills some extra conditions. As an example, many flat or deSitter
kink-metrics (also in 3+1 dimensions) have been studied \pcite{DeSitterkinks,
Dunn,Vasilic,PhysLetts}. At first sight, however, these solutions seem to
be in conflict with a well-established
approach: Any maximally extended multiply connected spacetime should occur as
a factor space of the simply connected {\em universal covering\/} solution;
but neither from Minkowski space nor from the universal covering of deSitter
space, kink solutions could be obtained in this way \pcite{Wolf,III}. This
apparent paradox is resolved by noting that --- even when requiring simple
connectedness --- there is no unique maximal (analytic) extension of a manifold
(such a warning has e.g.\ already been expressed in footnote 15 of
\pcite{GerochZit}): Take, for instance, Minkowski space, cut out a point and
construct the universal covering of this punctured plane, which now winds
around the removed point infinitely often in new layers
(cf.\ e.g.\ Fig.\ 5({\em d\/})).
Clearly, this solution is no longer geodesically complete at the removed
point but nevertheless maximally extended, since adding the point again would
yield a conical singularity. Identifying different layers of this
manifold, maximally extended $(k\ne2)$-kinks can be obtained; only the
corresponding 2-kink is extendible, since insertion of the point
restores Minkowski space.

Certainly, these latter kink solutions (and in fact all flat kinks or kinks of
constant curvature) are never geodesically complete.
On the other hand, there are plenty of complete kinks.
\footnote{A simple example is obtained from a metric of the form \re{EF})
  with one triple zero of $h(r)$ and $h(r)\sim r^{n\le2}$ for
  $r\to \pm\infty$ (e.g.\ $h(r)=r-\arctan r$). As shown in \pcite{III,II}
  the maximal extensions
  of this metric are geodesically complete kinks of arbitrary kink-number.
  If $h(r)\sim r^{n>2}$ these kinks are no longer geodesically complete but
  nevertheless inextendible, since the curvature diverges at the incomplete
  boundaries then (e.g.\ the solutions {\bf G4, R2} in \pcite{III,II}).}
In the case that they have a Killing symmetry, a complete classification has
been provided in \pcite{III}.
In Section 2 we will show by the example of the bare kink how
searching rigorously for maximal extensions and applying the factorization
method of \pcite{III} allows us not only to shed some light on the geometry
of these kinks but even to derive a whole bunch of related new ones.
The same concepts --- when applied to ``incomplete universal coverings'' like
the one mentioned above --- will also yield more interesting examples of
incomplete kinks. This is demonstrated for the flat case in Section 3 and
generalized to arbitrary metrics with a Killing field in Section 4. 
In Section 4 (as well as at the end of Section 2 and in Appendix A) we will
also provide explicit coordinate representations for the newly obtained kink
metrics.

Much of the interest in spacetimes with non-trivial kink-number centres
around such spacetimes which are locally solutions to the field equations of
some gravity model, i.e.\ of some appropriate gravity action. Since e.g.\
all solutions of generalized dilaton gravity models have a Killing
field \pcite{Banks,Kunstatter,I}, the scheme of the present paper
allows for the construction of kink spacetimes for any of these
models. But also conversely, for any given metric with a (local) Killing
symmetry, and hence also for all the kink metrics it gives rise to
(cf.\ Secs.\ 2 and 4), there is some gravity action for which the
metric solves the corresponding field equations \pcite{I}.
Let us thus briefly recollect some results about those metrics.
Here it is advisable to use a non-conformal gauge for the metric,
in contrast to what is useful on other occasions
such as in string theory, where the action is invariant under
rescalings of the metric by a conformal factor:
\footnote{The use of {\em non}\/-conformal gauges proves to be especially
  powerful in the presence of a Killing field. As such, gauges closely
  related to Eq.\ (\ref{EF}) below have been used with success in the
  literature \pcite{Banks,nonconf,KuSchw}. In particular we use this
  opportunity to gratefully acknowledge here the influence of W.\
  Kummer on our work. Bringing to our attention the success
  of non-conformal gauge conditions in a 2D gravity model \pcite{KuSchw},
  was essential for our interest in two-dimensional gravity theories,
  culminating finally in a series of papers on this subject.}
Any 2D metric with a (local) Killing field may be represented locally
in the generalized Eddington-Finkelstein (EF) form
\begin{equation}
  g = 2 {\rm d}r {\rm d}v + h(r) {\rm d}v^2
 \plabel{EF}
\end{equation} 
for some function $h$ (cf, e.g., \pcite{II}). The Killing field in these
coordinates is clearly $\6_v$, its length squared equals $h$, and the
curvature scalar is $R = h''$.
For a given metric the function $h$ in \re{EF}) is generically unique
up to an equivalence relation $h(r) \sim b^2h({r\0 b}+a)$, $a,b=const$.
Only for Minkowski and deSitter space this is not quite true, because they
have more than one Killing field (in fact three).
For instance, Minkowski space can be described by $h^{Mink}(r)=ar+b$, where
different choices of $h^{Mink}(r)$ may correspond to qualitatively different
Killing fields $\6_v$: $h^{Mink}(r)=b$ implies that $\6_v$ generates
translations (timelike, null, or spacelike, according to
$\mbox{sgn}\,b$), whereas for $h^{Mink}$ linear in $r$
($a\ne0$) the vector field $\partial_v$ generates boosts.
\footnote{In this respect the latter EF-coordinates, e.g.\
  $g=2{\rm d}r{\rm d}v+r{\rm d}v^2$, resemble the Rindler coordinates
  \pcite{Rindler}, $g=x^2{\rm d}t^2-{\rm d}x^2$
  (substitute $r=x^2/4$, $v=2(t-\ln x)$), where also $\partial_t$
  generates boosts. However, the EF-coordinates cover a considerably larger
  portion of Minkowski space (cf.\ Fig.\ 8({\em b\/})).}
Likewise, (anti-)deSitter space of curvature
$R$ is described by $h^{deS}(r)={R\0 2}(r+a)^2+b$, and again there are
three qualitatively different
Killing fields $\6_v$ according to $\mbox{sgn}\,b$ (cf.\ e.g.\
\pcite{III,II}).

If the metric $g$ is analytic, furthermore, then the function $h$ is
characteristic for the whole spacetime and there is a unique analytic
simply-connected
extension where the boundary (to be defined properly) is either complete or
a curvature singularity (called ``global'' in \pcite{III}; without such a
requirement the extension is not unique, as mentioned before).
An exposition of simple rules of how to obtain this
extension and its Penrose diagram from a given function $h$ may be found in
\pcite{II}. The multiply connected global solutions can be obtained by
factoring these universal coverings by discrete symmetry groups or,
equivalently, by cutting out some region of the universal covering and
gluing appropriately \pcite{III}. Whereas the first approach is favourable
for a concise classification, the second one is more straightforward
and shall be employed here.

\section{The bare kink, its Penrose diagram, and generalizations}

An in some sense prototypical example of a kink-metric is the ``bare kink''
\pcite{Dunn,Bare}
\begin{equation}
  g=-\cos2x\,{\rm d}t^2-2\sin2x\,{\rm d}t\,{\rm d}x+\cos2x\,{\rm d}x^2\,\,.
  \plabel{barekink}
\end{equation}
Its null-extremals are calculated easily to 
\begin{equation}
  {{\rm d}t\0 {\rm d}x}=-\tan\left(x\pm{\pi\0 4}\right)\,,
\end{equation}
and it is thus clear that the lightcone tilts with increasing $x$
(cf.\ Fig.\ 1({\em a\/})).
\begin{figure}[t]
 \begin{center}
 \leavevmode
 \epsfxsize 14cm \epsfbox{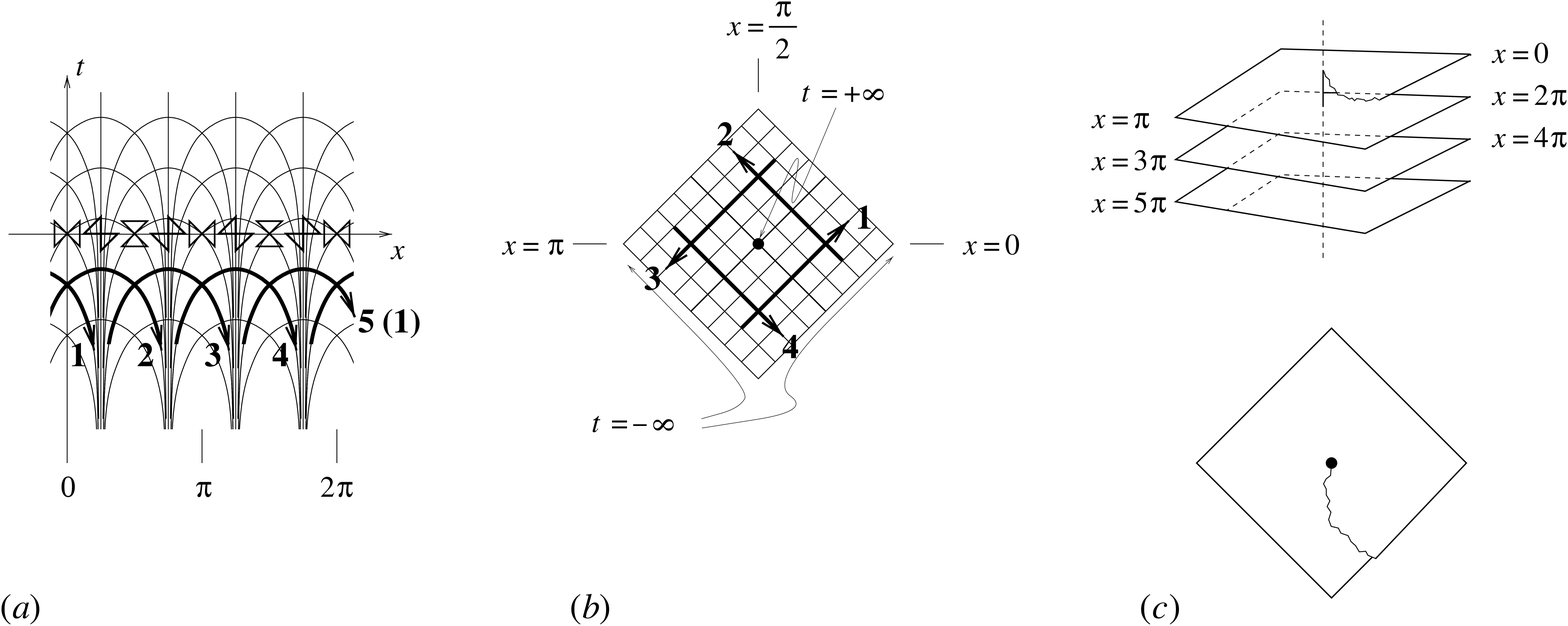}
 \end{center}
 \renewcommand{\baselinestretch}{.9}
 \small \normalsize
 \begin{quote}
 {\bf Figure 1:} {\small The kink-metric \re{barekink}) and its
   conformal derivates (like e.g.\ \re{flatkink})):
   ({\em a\/}) shows the lightcone structure and the null-extremals in
   the original $x,t$ coordinates. In a Penrose diagram of the corresponding
   coordinate patch ({\em b,c\/}) they play the role of polar coordinates,
   $x$ being the angle and $t$ some radial coordinate which goes
   $\rightarrow +\infty$ near the origin
   (for the extended Penrose diagram cf.\ Fig.\ 2).
   The 2-kink manifold ({\em b\/}) is obtained from ({\em a\/}) resp.\
   ({\em c\/}) by applying the identification $x\sim x+2\pi$, thereby mapping
   e.g.\ the null-extremal {\bf 5} onto {\bf 1}.}
 \end{quote}
\end{figure}
In order to obtain a Penrose diagram it is advisable to interpret $x$ and $t$
as polar coordinates,
\begin{equation}
  \mytilde x=e^{-t}\cos x\,,\quad \mytilde t=e^{-t}\sin x \, ,
  \plabel{polar}
\end{equation}
which brings \re{barekink}) immediately into conformally flat form:
\begin{equation}
  g={{\rm d}\mytilde t^2-{\rm d}\mytilde x^2 \0 \mytilde t^2+\mytilde x^2}\,
\end{equation}
Note that when applying \re{polar}) and thus also in the Penrose diagram
Fig.\ 1({\em b\/}) we have tacitly assumed that the identification
$x \sim x + 2\pi$ is made. If this is not desired, then there will occur
overlapping layers as displayed in Fig.\ 1({\em c\/}).
Obviously any kink-number $k$ can be obtained from this manifold by imposing
the identification $x \sim x + k\pi$. While for even kink-numbers this amounts
to identifying overlapping layers in ({\em c\/}), odd kink-numbers involve a
point-reflexion (inverting space and time).

This patch is, however, still incomplete! The central point $t=+\infty$ is
at an infinite affine distance, but the null-infinities at
$t\rightarrow-\infty$ are incomplete. A maximal extension can be
obtained using Eddington-Finkelstein coordinates and following the
recipe of \pcite{II},
\footnote{Substituting $x=r+{\pi\0 2}$, $t=\ln|\cos r-\sin
  r|-v$ into \re{barekink}), $g$ is easily brought into the form 
  \re{EF}) with $h(r)=\cos2r$.  This means, according to \pcite{II}, that the
  building block (as defined there) 
  is infinite, periodic, with non-degenerate horizons, and the
  maximal extension (if identifying overlapping layers) is the chessboard-like
  arrangement of Fig.\ 2.}
or, as is even simpler in the present context, by applying the
transformation $\mytilde x \pm \mytilde t=\tan(\hat x \pm \hat t)$,
which yields\footnote{A similar form of the
metric has also been found in \pcite{globkink}.}
\begin{equation}
  g={{\rm d}\hat t^2-{\rm d}\hat x^2\0\sin^2\hat t+\sin^2\hat x} \,\,.
  \plabel{barekinkglobal}
\end{equation}
The scalar curvature for this metric,
\begin{equation}
  R=4{\cos 2\hat t - \cos 2\hat x \0 - 2 + \cos 2\hat t + \cos 2\hat x}\,,
\end{equation}
ranges between $R=-4$ at $\hat t=n\p$ (white regions in Fig.\ 2)
and $R=4$ at $\hat x=n\p$ (dark shaded regions).
At $\hat t, \hat x=n\p$ the metric becomes
singular, and these points (full circles in Fig.\ 2, where the
Killing trajectories meet) are at an infinite distance. Thus to obtain the
universal covering the overlapping sectors after surrounding those points
should {\em not\/} be identified (cf.\ Fig.\ 1({\em c\/})). Disregarding this
multilayered structure (after all, the 2-kink is single-layered!),
\re{barekinkglobal}) provides a global chart for the bare kink. 
\begin{figure}[htb]
 \begin{center}
 \leavevmode
 \epsfxsize 9cm \epsfbox{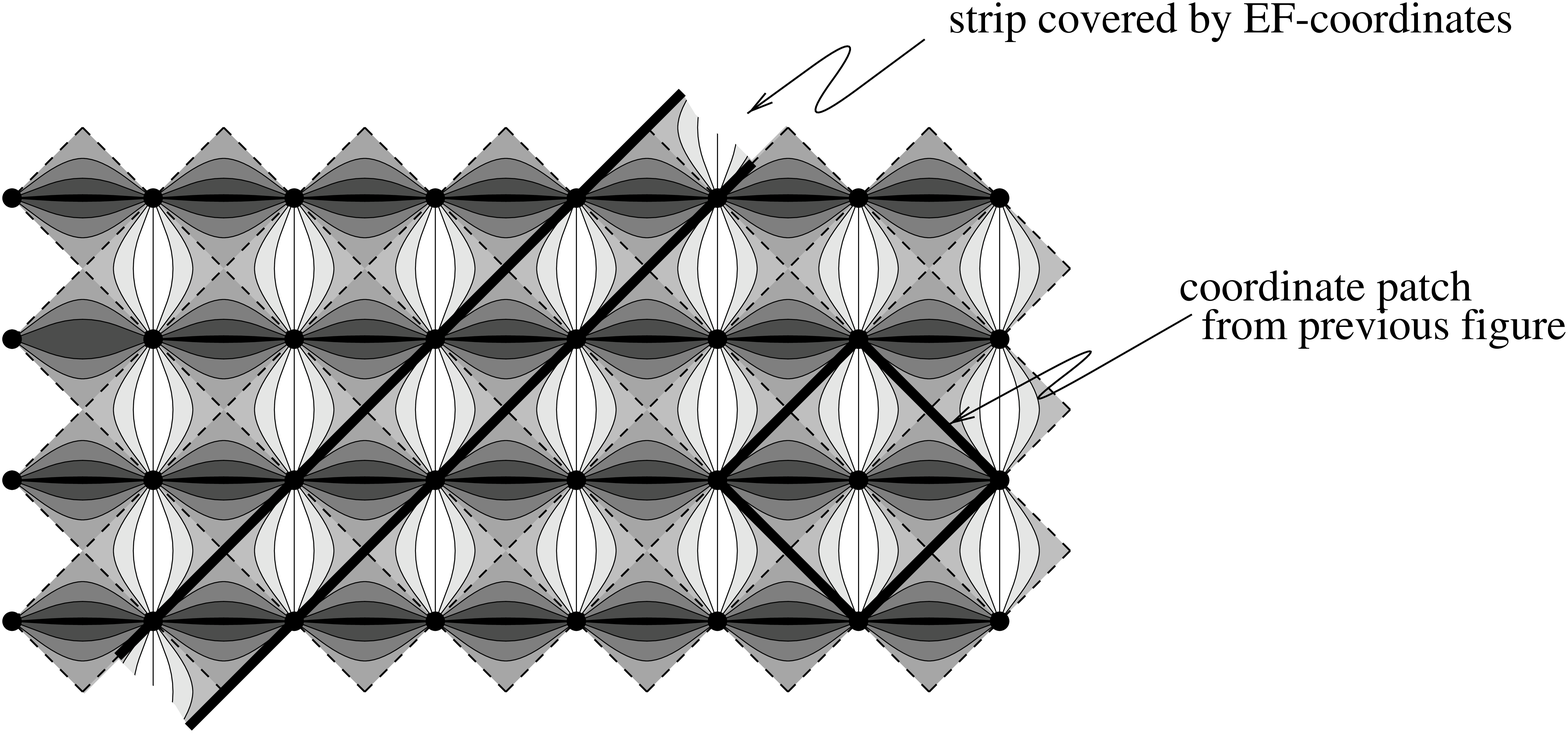}
 \end{center}
 \renewcommand{\baselinestretch}{.9}
 \small \normalsize
 \begin{quote}
 {\bf Figure 2:} {\small A maximal extension of the bare kink, to be
   infinitely extended. The shading corresponds to the curvature ranging over
   $-4\le R \le 4$, the thin $R=const$-lines being at the same time Killing
   trajectories for the field $\6_t$. The coordinate patch \re{barekink})
   (with the identification $x\sim x+2\pi$) is shown, and also the patch
   covered by EF-coordinates \re{EF}) with $h(r)=\cos 2r$.
   The universal covering would be obtained, if at every full circle
   (indicating points at an infinite distance) the manifold were continued
   into a new overlapping layer as sketched in Fig.\ 1({\em c\/}).}
 \end{quote}
\end{figure}

What can be learned from this representation, taking into account the
method of \pcite{III}?
First, even when starting from a cylindrical kink, the maximal extension
need not be a cylinder. Already the manifold displayed in Fig.\ 2
provides a counterexample, but many more may be found:
for instance, gluing together opposite faces of the
diamond-shaped coordinate patch yields a torus with two holes,
Fig.\ 3({\em a\/}), and likewise ({\em b\/}) yields a torus
with one hole.
\begin{figure}[htb]
 \begin{center}
 \leavevmode
 \epsfxsize 12cm \epsfbox{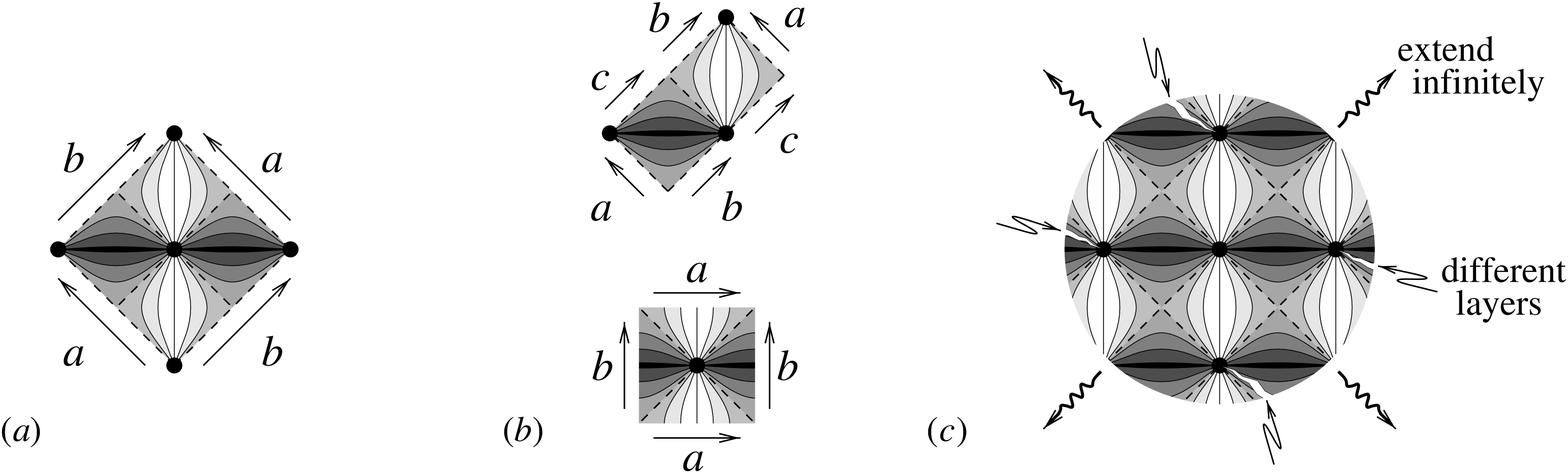}
 \end{center}
 \renewcommand{\baselinestretch}{.9}
 \small \normalsize
 \begin{quote}
 {\bf Figure 3:} {\small Some further maximal extensions of the
   bare kink. ({\em a\/}) yields a torus with two holes, whereas ({\em b\/})
   shows two equivalent constructions of a torus with one hole. It is also
   possible to obtain a globally cylindrical extension of the original kink,
   by repeating the construction of the universal covering, i.e., at any
   complete point except for the very first one the manifold has to be
   extended into different overlapping layers ({\em c\/}).}
 \end{quote}
\end{figure}
On the other hand, proper topological cylinders will always be found among the
covering solutions: just start from the original punctured diamond but then
proceed in the same way as if constructing the universal covering, i.e.,
whenever surrounding a complete point (full circle) start with a new
overlapping layer (Fig.\ 3({\em c\/})).
Topologically, this infinitely branching extension cannot be distinguished
from a cylinder, though, admittedly, the causal structure near the exterior
``frazzled'' boundary will be rather involved.
And many more cylindrical kinks can be obtained in a similar manner:
Start from {\em any\/} non-contractible closed ribbon like $A$ or $C$ in
Fig.\ 4\ ($A$ giving rise to the kink described previously)
resp.\ any open ribbon covering (part of) the manifold like $D$ with
its short edges identified.
Whenever the overall number of tilts of the lightcone when going along the
ribbon does not vanish, this is a kink manifold (cf.\ the introduction)
and proceeding as before a maximal cylindrical extension may be constructed.
It is straightforward to give an explicit atlas for these manifolds, using
EF-coordinates \re{EF}) for straight segments and, e.g., \re{barekink})
for the bent ones.
\begin{figure}[htb]
 \begin{center}
 \leavevmode
 \epsfxsize 10cm \epsfbox{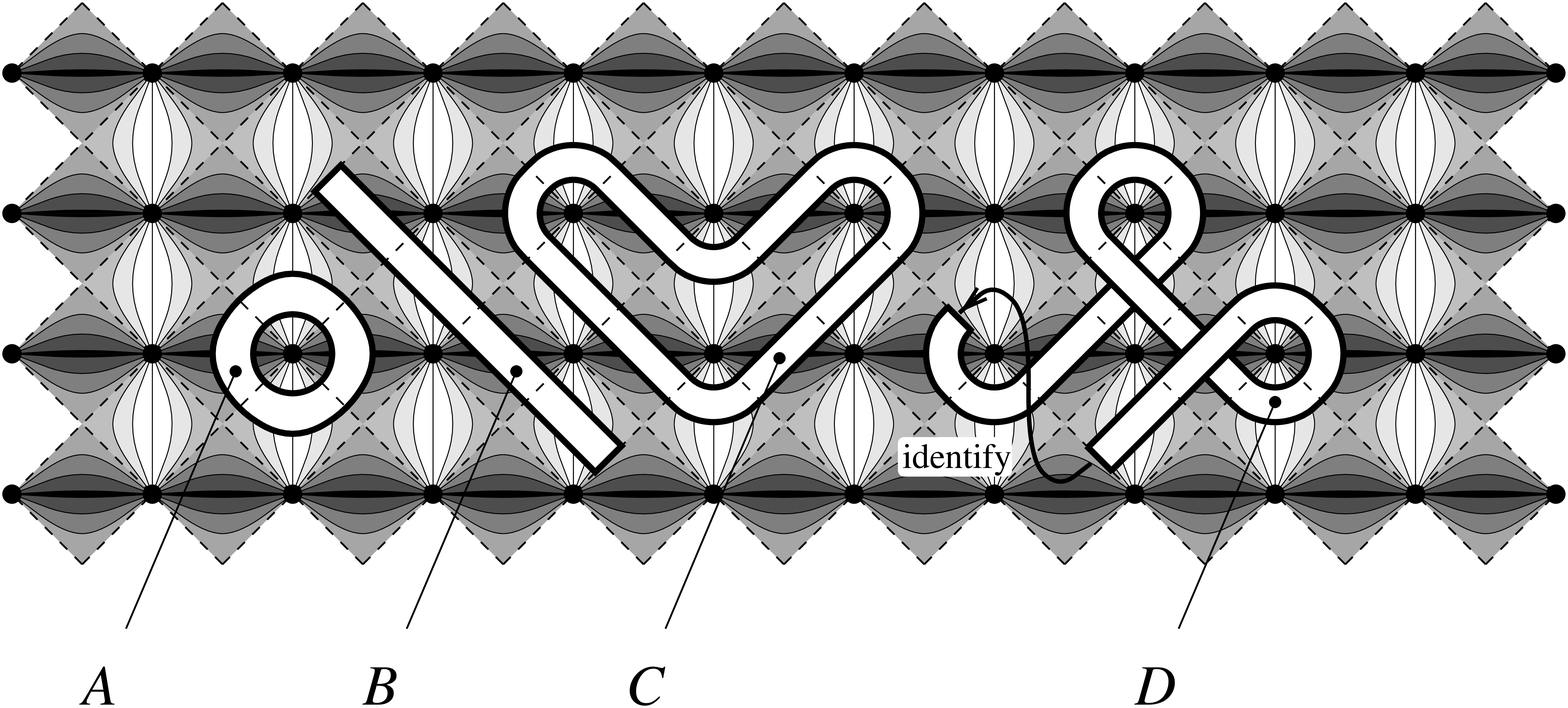}
 \end{center}
 \renewcommand{\baselinestretch}{.9}
 \small \normalsize
 \begin{quote}
 {\bf Figure 4:} {\small Further manifolds derived from the bare
   kink. Whereas $A$ and $C$ are 2-kinks (cylindrical, if extended suitably),
   $D$ with the short edges identified depicts a 4-kink. Identifying the
   opposite faces of strip $B$ yields an incomplete ``Taub-NUT''
   torus resembling in this respect the Misner torus, which is obtained from
   $A$ by gluing the inner to the outer boundary. Similar incomplete tori may
   be constructed from $C$ and $D$ (again identifying inside and outside).}
 \end{quote}
\end{figure}

Second, and perhaps even more surprising: a continuous family of locally
equivalent but globally different kinks can be obtained. Note that
(in the spirit of \pcite{III}) already the original $k$-kink \re{barekink})
is obtained by identifying overlapping sectors of the universal covering
(i.e.\ gluing $x\sim x+k\p$). However, since the bare kink has a Killing
symmetry ($\6_t$ in coordinates \re{barekink})), one may apply a Killing
transformation prior to the gluing. This amounts to an identification
$(x,t)\sim(x+k\p,t+\p k\a)$ in \re{barekink}), i.e.\ in Fig.\
1({\em a}) a vertical shift of length $k\p\a$ is applied before
gluing. Alternatively, one could as well substitute $t\rightarrow t+\a x$
into \re{barekink}),
while sticking to the original identification $(x,t)\sim(x+k\p,t)$.
The metric for these ``not-so-bare'' kinks can then be written as
\begin{equation}
  g=-\cos2x\,{\rm d}t^2-2(\sin2x+\a\cos2x)\,{\rm d}t\,{\rm d}x
    +\left((1-\a^2)\cos2x-2\a\sin2x\right){\rm d}x^2\,\,.
  \plabel{notsobarekink}
\end{equation}
For a proof that these solutions are not isometric for different $\a$ and
for a geometrical characterization of this parameter cf.\ \pcite{III}.
Of course, the same construction applies also to the more elaborate
kinks discussed before (like those obtained from $C$, $D$ of Fig.\
4), introducing a
continuous parameter also there, and furthermore to non-cylindrical
extensions, where each generator of $\pi_1(\CM)$ acquires a parameter.

According to \pcite{III} the above constructions (i.e.\ gluing isometric
sectors of the universal covering, thereby introducing a continuous
parameter for each generator of $\pi_1(\CM)$) exhaust the complete manifolds
obtained from the bare kink metric. However, there are further incomplete
yet inextendible manifolds: One such example is the Misner-torus (see the
following paragraph) which could only be extended when abandoning the
Hausdorff property. But also the constructions of section 4 can be applied
(leading to \re{kink}) with $h(r)=\cos 2r$), in which case smoothness
prohibits a further extension of the resulting kinks.

Finally, we shortly touch the case of another manifold which can be derived
from the bare kink metric: the Misner-torus \pcite{Bare,Mis}. It is obtained
by also wrapping up periodically the coordinate $t$ in \re{barekink}),
$t\sim t+\o$ (besides $x\sim x+2\p$), or equivalently from the annulus $A$
shown in Fig.\ 4\ by identifying the inner and
outer boundary along the Killing trajectories (thin lines).
This metric, although well-behaved everywhere on a compact manifold,
is incomplete near the Killing horizons (dashed lines), the
singularity being of the Taub-NUT type \pcite{Mis,Geroch}. Given the
extended bare-kink manifold, however, many more tori of that type
can be obtained. For instance, gluing opposite faces of the strip $B$
yields another torus with similar pathological completeness
properties.
\footnote{The tori obtained from $A$ and $B$ are indeed inequivalent:
  whereas in $B$ there are complete null-extremals (the ones running
  alongside of the displayed strip), this is not the case in Misner's
  example $A$.}
Explicit coordinates for the torus $B$ are easily obtained from the
EF-coordinates $ g = 2 {\rm d}r {\rm d}v + \cos2r {\rm d}v^2 $ by
identifying $r\sim r+2\p$ as well as $v\sim v+\o$.
But even more exotic specimens may be constructed, e.g.\ starting
from $C$ or $D$ and gluing again the two boundaries.
And certainly the parameter $\a$ from \re{notsobarekink}) can be introduced
also here; together with $\omega$ this yields the two parameters for
the two generators of $\pi_1(\mbox{torus}) = \Z^2$.

\section{Flat kinks} 

Adding a simple conformal factor to the metric \re{barekink}) yields quite a
different example. The resulting metric
\begin{equation}
  g=e^{-2t}\left(-\cos2x\,{\rm d}t^2-2\sin2x\,{\rm d}t\,{\rm d}x +
    \cos2x\,{\rm d}x^2\right)
  \plabel{flatkink}
\end{equation}
has the same tilting-lightcone structure as \re{barekink})
and is thus also a kink-metric. Calculating the curvature of this metric
shows, however, that it is actually flat: $R\equiv0$.

Again the transformation \re{polar}) clarifies the situation 
(as before we choose the 2-kink version of \re{flatkink}),
i.e.\ identify $x\sim x+2\p$):
This time it leads to 
\begin{equation}
  g={\rm d}\mytilde t^2-{\rm d}\mytilde x^2 \,\plabel{Mink},
\end{equation}
so this is nothing but flat Minkowski space in polar coordinates, the
origin being removed.  In contrast to \re{barekink}) the metric
\re{flatkink}) is incomplete at the origin; it has a hole which can
(and thus should) be filled by inserting a point, leaving ordinary
Minkowski space without any kink.
In retrospect thus \re{flatkink}) is a rather blunt construction of a
kink. In fact, {\em any\/} nice no-kink manifold gives rise to many
such kinks simply by cutting holes into it: Since the lightcone tilts
upside-down twice when surrounding the hole, this is a 2-kink
solution. Taking covering spaces arbitrary even kink-numbers can be obtained
(odd kink-numbers may occur, if there is a point-reflexion symmetry).
Of course, these $k\ne2$-kinks are no longer extendible, because at the
inserted point there would occur a conical singularity (branch point)
and the extension could not be smooth. Introducing polar coordinates
\re{polar}) it is straightforward to write down explicit charts for these
kinks (identifying $x\sim x+k\p$ again).
Nevertheless, this construction seems rather artificial.

\begin{figure}[htb]
 \begin{center}
 \leavevmode
 \epsfxsize 14cm \epsfbox{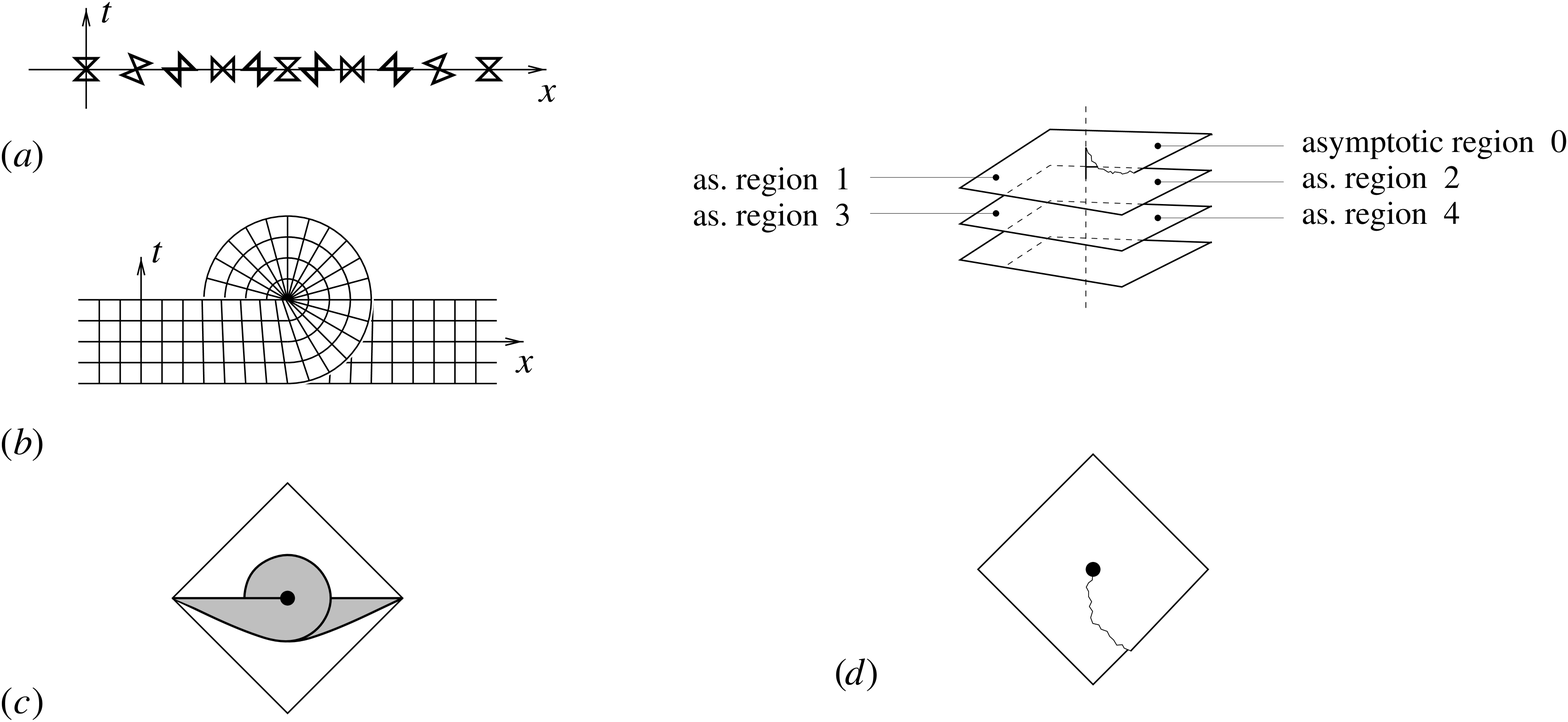}
 \end{center}
 \renewcommand{\baselinestretch}{.9}
 \small \normalsize
 \begin{quote}
 {\bf Figure 5:} {\small Flat 2-kink of topology $\R\times\R$:
   ({\em a\/}) shows the lightcone structure in the original $x,t$-coordinates,
   ({\em b\/}) an ``embedding'' (allowing for overlapping layers) of this
   patch into Minkowski space, and ({\em c\/}) the corresponding Penrose
   diagram. Its maximal extension, however, contains infinitely many
   asymptotic regions ({\em d\/}) and thus subsets of arbitrary kink-number.
   Note that ({\em d\/}) represents an incomplete but inextendible simply
   connected flat manifold different from Minkowski space.}
 \end{quote}
\end{figure}
Note that in a similar manner ``kinks'' of topology $\R\times\R$ may be
obtained. This refers to coordinates $x,t$ where the lightcone tilts a couple
of times when going from one asymptotic region, $x\to-\infty$, to the other,
$x\to\infty$ (cf.\ Fig.\ 5({\em a\/}) for kink-number 2).
However, in 2D the maximal extension of such a kink usually has an infinite
series of potential asymptotic regions (cf.\ Fig.\ 5({\em d\/}))
and certainly one could as well choose coordinates which wrap around the
origin more often before settling down in an asymptotic region. Thus
$\R\times\R$-kinks of arbitrary kink-number can be obtained
(in contrast to the conclusion of \pcite{Vasilic}).
On the other hand, this kink-number merely characterizes the coordinates,
not the (extended) spacetime itself, which is always of the form ({\em d\/})
(or a cylindrical kink of sufficiently large kink-number);
\footnote{This is a 2D artifact, however: In higher dimensions the
  asymptotic region is usually topologically sufficiently non-trivial to allow
  for ``genuine'' kink-numbers {\em within\/} this region (if connected), or
  within its connected components, respectively.}
for these reasons we dismiss this topic and return to cylindrical kinks again.

It has been pointed out already that there are no geodesically complete
flat kinks. The kinks \re{flatkink}), on the other hand, are at least
inextendible (due to the conical singularity) except for the $k=2$ case,
which in some sense reveals their construction (cutting a hole).
However, in the presence of a Killing field --- and in the case of
Minkowski space there are three independent Killing fields ---
we can do more than just cutting out points or regions. This will allow
us to construct continuous families of inequivalent flat kinks, which
are inextendible even in the $k=2$ case. Moreover, the procedure
allows for a straightforward generalization to any metric with a
Killing field, as will be shown in section 4.

Start from Minkowski space, but instead of only removing the
origin cut out a whole wedge (cf.\ Fig.\ 6({\it a\/})).
By means of a Lorentz boost the two edges of the wedge can be mapped onto
one another, and we use this boost to glue the remaining patch together
(note that also the tangents at these edges must be mapped with the
tangential map of this boost). 
\begin{figure}[htb]
 \begin{center}
 \leavevmode
 \epsfxsize 9cm \epsfbox{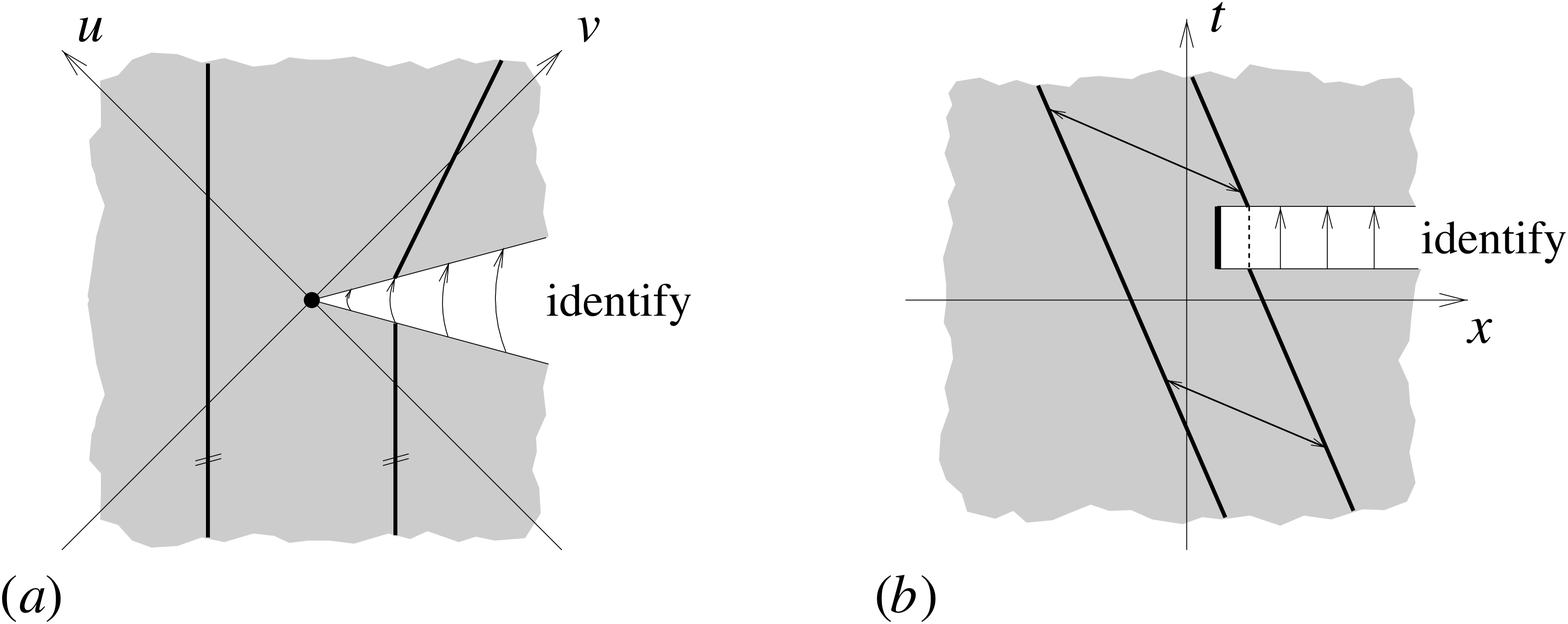}
 \end{center}
 \renewcommand{\baselinestretch}{.9}
 \small \normalsize
 \begin{quote}
 {\bf Figure 6:} {\small ({\it a\/}) Minkowski kink with non-trivial
   holonomy. This space can be obtained by removing a wedge from flat Minkowski
   space and gluing together the corresponding boundary lines by a boost.
   Due to this construction two extremals which are parallel on one side of the
   origin are mutually boosted on the other side (cf.\ bold lines).
   Thus the holonomy is non-trivial (surrounding the origin yields a boosted
   frame), and at the origin there would occur a conical singularity.
   ({\it b\/}) Another Minkowski kink; it has trivial holonomy but the
   distance of parallels passing the hole changes (for a possible maximal
   extension cf.\ Fig.\ 12).}
 \end{quote}
\end{figure}
Clearly such a space is everywhere flat (except at the origin, which is
considered not to belong to the manifold) but it has non-trivial holonomy.
For instance, two timelike extremals which are parallel ``before'' passing the
origin at different sides will be mutually boosted afterwards (bold lines in
Fig.\ 6({\em a})). 
Also, due to the non-trivial holonomy the origin
can no longer be inserted (as was possible for the trivial Minkowski 2-kink
\re{flatkink})), since there would occur a conical singularity. Thus a
continuous one-parameter family of 2-kinks is obtained, labelled by their
holonomy (i.e.\ boost-parameter, or angle of the removed wedge). Changing
the sign of the boost-parameter corresponds to {\em inserting\/} a
wedge or, equivalently, to removing a ``timelike'' wedge.
Of course, taking covering solutions (with an additional factoring
by a point reflection for odd $k$)
arbitrary kink-numbers $k$ can be obtained, and for a given $k$ they are
again characterized by the additional boost-parameter.

Due to the non-trivial holonomy around the origin, there is certainly
no global chart on the punctured plane where $g$ takes the standard
Minkowski form $g=2{\rm d}u{\rm d}v$ (or \re{Mink})). But there is not
even a (globally smooth) conformal chart in this case. This may be seen as
follows: A metric in conformal gauge is flat, iff the conformal factor
equals a product of two functions of the lightcone coordinates $u$ and
$v$: $g=f_1(u)f_2(v)\,{\rm d}u{\rm d}v$. But any such metric can be extended
smoothly into the origin and is flat there. This contradicts the
assumption of a non-trivial holonomy (which necessarily entails a conical
singularity at the origin). Still, a smooth conformal coordinate system may
be found on a cylinder which only misses an arbitrarily small square around
the puncture. This chart will be provided in Appendix A,
cf.\ Eq.\  \re{boostkink}).
We remark here only that in the limit of shrinking the square to the origin,
the conformal factor given in Appendix A becomes distributional
(non-smooth even outside the origin).

In the above example we have chosen the boosts centered at the origin
as Killing symmetry. However, Minkowski space also exhibits translation
symmetries. An analogous construction can be applied in this case, too,
with the following geometrical interpretation:
cut out a whole slit (in direction of the chosen translation),
remove the strip on one side of the slit, and glue together the resulting
faces (cf.\ Fig.\ 6({\it b\/})). This manifold has now trivial
holonomy, but the metric distance of two generic parallels passing the hole
changes. Thus the manifold is so badly distorted that it cannot be completed
to ordinary Minkowski space, either. Also for this case smooth conformal
coordinates on a cylindrical region can be given, \re{noboostkink}), but
even analytic coordinates may be constructed, as will be demonstrated
in the following section.
However, in contrast to the former example this space can still be smoothly
extended further, leaving behind only branch points. This is shown in
detail in Appendix B, cf.\ Fig.\ 12.

\section{Kinks from arbitrary metrics and explicit coordinates}

In the previous section several examples of flat kinks have been constructed.
Clearly, since any 2D metric is locally conformally flat, some of these
methods will be applicable also to non-flat metrics: For instance, to obtain
a $2n$-kink from an arbitrary metric choose locally a conformal
gauge, change to polar coordinates \re{polar}) and identify $x\sim x+2\pi n$,
leading to \re{flatkink}) with another conformal factor. Again, however,
only the resulting $k\ne 2$-kinks (i.e.\ $n\ne1$) will be inextendible.
The more elaborate constructions of Fig.\ 6, on the other hand,
may be transferred only if the metric in question has a Killing symmetry.
(Note that this is e.g.\ the case for practically  all 2D vacuum gravity
models, cf.\ \pcite{I}). Using, for instance, local EF-coordinates \re{EF}),
the construction of Fig.\ 6({\em b\/}) can be applied literally,
with $(r,v)$ taking the role of $(x,t)$, since the Killing field $\6_v$
generates vertical translations then.

Remarkably, it is even possible to obtain an explicitly analytic chart for
such a kink (again at the cost of enlarging the hole slightly).
Let $(r,v)$ denote the original $\mbox{(EF-)}$\hspace{0cm}coordinates,
$(x,t)$ the new kink-coordinates,
and, without loss of generality, let the ``hole'' be centered at the origin,
$r,v=0$. Then the desired transformation is 
\begin{equation}
    r = t \cos nx \; , \quad 
    v = t \sin nx - \ell x \; ,
    \quad n \in \N \, ,    
  \plabel{koord}  
\end{equation}
with, e.g., $0\le x<2\p$. If the term to the right
in the expression for $v$ were absent, this would merely
be the formula for polar coordinates. The extra term accomplishes the
``cutting'' by producing a shift of length $l=2\p\ell$ in $v$-direction at
each full turn-around of the ``angle''-variable $x$. This is illustrated in
Figs.\ 7\ for $\ell>0$, $n=1$.
 
\begin{figure}[htb]
 \begin{center}
 \leavevmode
 \epsfxsize 12cm \epsfbox{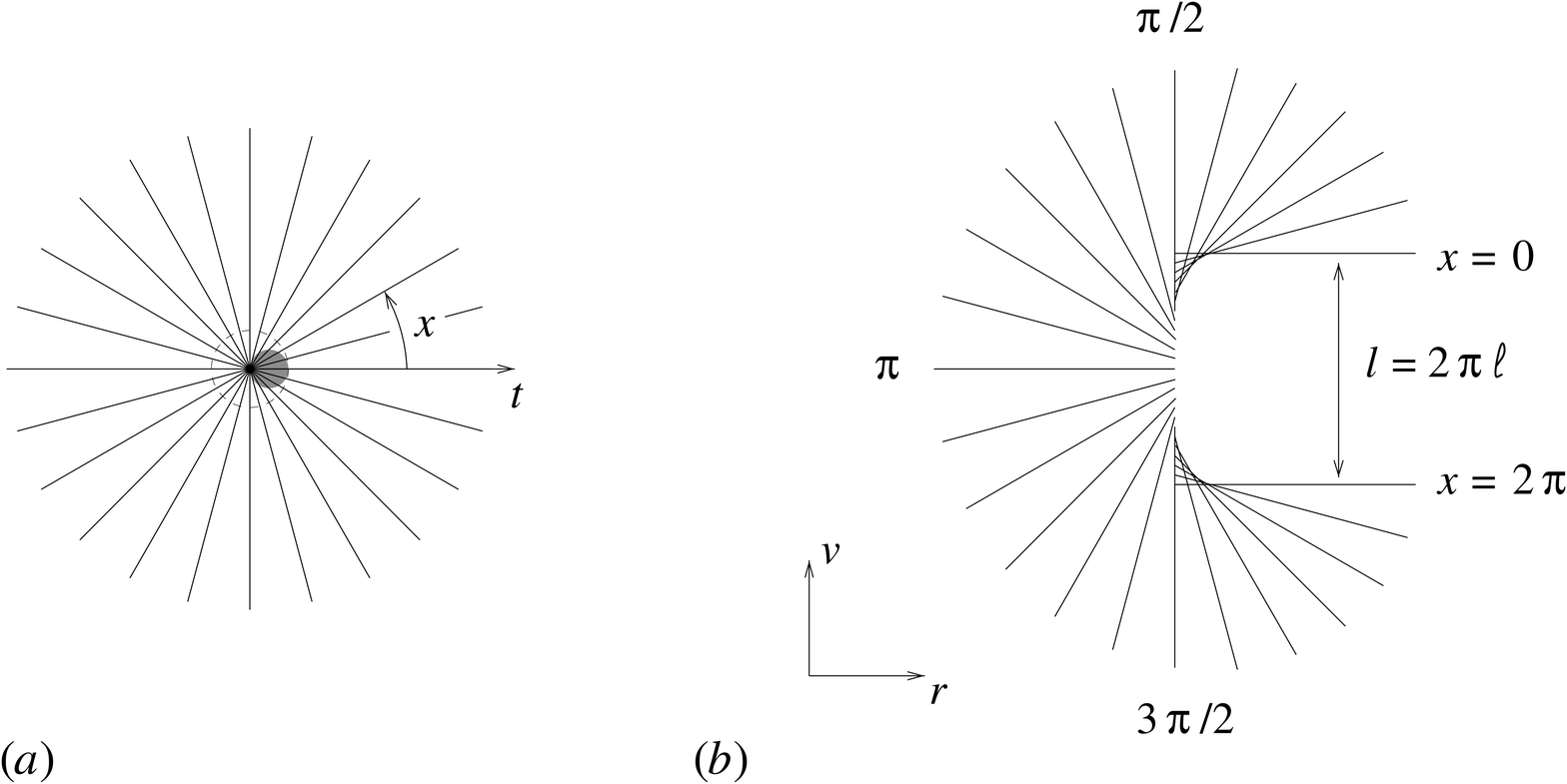}
 \end{center}
 \renewcommand{\baselinestretch}{.9}
 \small \normalsize
 \begin{quote}
 {\bf Figure 7:} {\small The cutting procedure \re{koord}) for $n=1$
   (2-kink). The patch ({\it a\/}) carries polar coordinates. The radii are
   mapped onto the corresponding lines in the diagram ({\it b\/}), where the
   EF-metric \re{EF}) lives. The pullback of this metric is the desired
   kink-metric. It is well-defined everywhere in ({\it a\/}) except for the
   small shaded
   disk indicated there. While the transformation as such is discontinuous at
   $x=0$, $2\pi$, \dots, this does not matter for the resulting metric,
   since everything is $v$-independent.}
 \end{quote}
\end{figure}
In the left diagram ({\it a\/}) $x,t$ are polar coordinates (to be
interpreted as cylinder coordinates for the kink spacetime, finally).
The radii $x=const$ are mapped in a
one-to-one fashion onto the rays in the right diagram ({\it b\/}), where
$r,v$ are Cartesian coordinates (carrying the EF-metric).  Of course this
transformation breaks down near the origin, as is seen by the intersecting
rays in ({\it b\/}). If the corresponding region (eccentric shaded disk)
is excluded from ({\it a\/}),
e.g.\ by the stronger restriction $t>|\ell|/n$ (broken circle), then we are
left with an ``annulus'' on which \re{koord}) is a local diffeomorphism.
(The jump from $x=2\p$ to $x=0$ does not raise any problems, since by
its $v$-homogeneity the EF-metric $g$ remains smooth.)

Inserting \re{koord}) into the EF-metric \re{EF}) leads to the desired
kink-metric. Since the expressions are rather ugly, we will only
write them down in terms of a zweibein. Note that via $g=2e^+e^-$ 
the metric \re{EF}) can be obtained from 
$e^+ = {\rm d}v  , \; e^- = {\rm d}r + \mbox{$\2$} h(r) {\rm d}v$.   
Transforming this zweibein into the new $x,t$-coordinates yields
\begin{eqnarray}
  e^+ &=& \sin nx\,{\rm d}t +\left[nt\cos nx-\ell\right]{\rm d}x \;
                                                              ,\nonumber\\
  e^- &=& {\rm d}\left[t \cos nx \right] + 
       {h\left(t\cos nx \right)\0 2}e^+
                               \; , \quad n \in \N \; .
  \plabel{kink}
\end{eqnarray}
As the expression \re{kink}) is $2\pi$-periodic in $x$, $g=2e^+e^-$
provides a metric on a cylindrical spacetime. The non-degeneracy
of $g$ is guaranteed (only) for values $t > |\ell|/n$,
\footnote{If preferred, one can of course reparametrize the
  $t$-coordinate, e.g.\ $t \rightarrow |\ell|/n+e^t$, so as to
  obtain a metric defined for all (coordinate) values of $t$.}
since
\begin{equation}
  e^-\wedge e^+ = (nt - \ell \cos nx)\,{\rm d}t\wedge {\rm d}x  \; .
  \plabel{funk}
\end{equation}
It is a $2n$-kink metric, i.e. moving along any circle $t =const > |\ell|/n$,
the lightcone makes $2n$ half-turns or $n$ full turns. This is evident by
construction, but may be verified also explicitly: Obviously
$e^+_\m = (\sin nx,nt \cos nx-\ell)$ winds around the origin $(0,0)$ $n$~times
when going along a $t=const$ loop. Due to its linear independence,
Eq.\ \re{funk}), $e^-_\m$ is doomed to follow this tilting movement. This
concludes the proof, since the null-directions are determined precisely by
$e^\pm$.

As already pointed out, the function $h$ in \re{kink}) is completely
arbitrary and can be chosen to fulfill the e.o.m.\ of any desired gravity
model \pcite{I}.
For instance, if $h$  is taken to be the function of
deSitter gravity, e.g.\ $h(r)=-1-r^2$, then we arrive exactly at the
deSitter-kinks of \pcite{PhysLetts}. These solutions are perfectly smooth
everywhere (in contrast to those of \pcite{Dunn}) but still non-trivial 
(i.e.\ not extendible to the global deSitter space).
\footnote{Our present approach allows us to correct a small mistake
  in \pcite{PhysLetts}: At that time it was believed
  that the spacetime described
  by \re{kink}) (with $h(r)=-1-r^2$) contains closed timelike curves. It is
  obvious from the present analysis that this is not the case. Quite on the
  contrary, for a cylindrical solution \re{kink}) there is {\em not a
  single\/} closed loop with a definite sign of ${\rm d}s^2$. As
  a consequence there is no foliation of the cylinder into spacelike leaves
  $\S \sim S^1$ (a Hamiltonian formulation, however, may still be defined,
  as demonstrated in \pcite{PhysLetts,IV}).} 

Another instructive example is obtained when taking $h(r)\propto r$.
As discussed in the introduction this yields flat space, but with the Killing
field $\partial_v$ describing boosts. Thus, when inserting
$h(r)\propto r$ into \re{kink}) one would expect to recover the kinks of
Fig.\ 6({\em a\/}). This is only partially true, however:
Substituting e.g.\ $r=-\frac{UV}{2}, v=2\ln\frac{V}{2}$  into the metric
$g=2{\rm d}r{\rm d}v+r{\rm d}v^2$ yields  $g=2{\rm d}U{\rm d}V$ with $V>0$.
Thus, the above EF-coordinates underlying Fig.\ 8({\em a\/}) cover
only half of Minkowski space, namely the dark shaded region of ({\em b\/}).
Subsequently, one can certainly replace the removed strip by a wedge (of
the same ``boost width'') as shown in ({\em c\/}). This patch coincides with
the one of Fig.\ 6({\em a\/}), displayed for comparison in
Fig.\ 8({\em d\/}) again, only on an annular (i.e.\ cylindrical)
region remote from the origin (e.g.\ outside the dashed ellipse).
This also shows once more that a not maximally
extended kink can have quite different extensions (note that the patch
({\em c\/}) has to be extended further, as shown in Appendix B,
Fig.\ 12).
\begin{figure}[htb]
 \begin{center}
 \leavevmode
 \epsfxsize 15cm \epsfbox{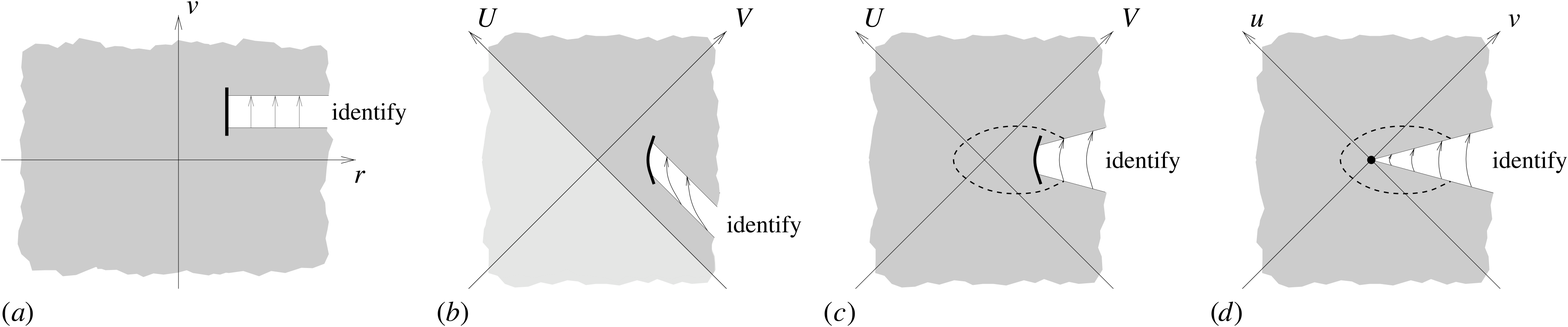}
 \end{center}
 \renewcommand{\baselinestretch}{.9}
 \small \normalsize
 \begin{quote}
 {\bf Figure 8:} {\small Relationship between the kinks \re{kink})
   and Fig.\ 6({\it a\/}): The EF-coordinates ({\it a\/}) with
   $h^{Mink}(r)\propto r$ cover half of Minkowski space ({\it b\/}).
   Instead of removing a strip bounded by null-lines, one can remove a wedge
   ({\it c\/}). On a cylindrical region outside the dashed ellipse this
   manifold coincides with the Fig.\ 6({\it a\/}) ``boost''-kink
   ({\it d\/}).}
 \end{quote}
\end{figure}

How exhaustive is the construction scheme described above? First of all, it has
to be pointed out that we have only shown the simplest examples. In general
{\em any\/} cylindrical {\em limited\/} covering
\footnote{In contrast to the familiar (unlimited) covering manifolds, a
  limited covering may have (ideal) boundaries which do not correspond to
  boundaries of the underlying base manifold. Especially, any open subset of
  a manifold is a limited covering.}
of the original spacetime or,
worse, any factor space thereof with a nonzero number of tilts of the lightcone
may be addressed as kink (remember e.g.\ the manifold $D$ of Fig.\
4). It is certainly tempting to assume that the maximally
extended universal covering of any such kink is a branched covering of the
unique ``global'' universal covering (as constructed in \pcite{II}),
although we could not prove this assertion. Any inextendible kink solution
could then be obtained as a factor space from the respective branched covering
and a classification would amount to specifying the position of the branch
points (up to symmetry transformations, of course) and finding the conjugacy
classes of freely and properly discontinuously acting symmetry subgroups for
this manifold \pcite{Wolf,III}. For globally cylindrical
kinks, furthermore, only the subgroups isomorphic to \Z\ are relevant.
However, even if this approach is feasible it will sometimes obscure
geometric features: Note that generically the presence of a branch point
breaks the Killing symmetry (unless the Killing vector vanishes at that
point, as is the case in Fig.\ 6({\em a\/})). Consequently, when describing
the kink of Fig.\ 6({\em b\/}) as a factor space of its maximally extended
universal covering (cf.\ Appendix B), the continuous parameter ``translation
width'' does not emerge from a Killing symmetry during the factorization, but
it is already encoded in the spacing of the branch points in the universal
covering.

We conclude this section with two further constructions which have not been
made explicit so far. First, if there are symmetries not generated by a
Killing field, then further discrete parameters may be introduced.
This is e.g.\ the case for the Reissner-Nordstr\"om solution, which is an
infinite periodic repetition of one patch: One could make a long vertical slit
through a number of patches, then remove a few patches on one side 
and glue together again, cf.\ Fig.\ 9. The resulting kink is thus
characterized by a Killing parameter {\em and\/} the patch number.
\begin{figure}[htb]
 \begin{center}
 \leavevmode
 \epsfxsize 3cm \epsfbox{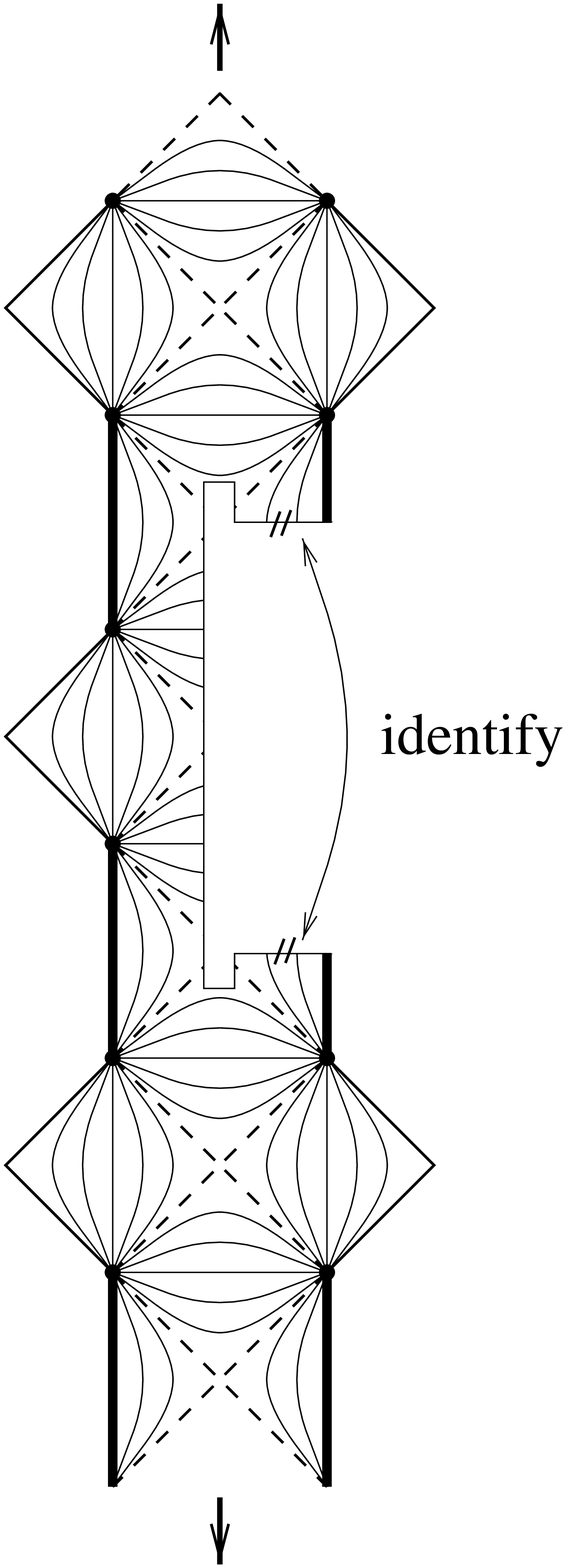}
 \end{center}
 \renewcommand{\baselinestretch}{.9}
 \small \normalsize
 \begin{quote}
 {\bf Figure 9:} {\small Reissner-Nordstr\"om kink.}
 \end{quote}
\end{figure}
Certainly, also this kink can be extended in a similar manner as the one of
Fig.\ 6({\em b\/}), cf.\ Appendix B.
Second, let us note that such surgery is obviously not restricted to
cylindrical solutions (i.e.\ one hole only), but within any solution one can
cut any number of holes, each giving rise to a kink-number $k$, a continuous
parameter for the gluing (if there is a Killing symmetry), further ones for
the relative position of the respective new hole, and perhaps
some discrete parameters. And applying more advanced gluing techniques
(like making slits between branch points and sewing the overlapping layers
cross-wise) it is even possible to obtain surfaces of higher genus.
\footnote{This is also well-known from complex analysis, where e.g.\ the
  Riemann surface of the function $\sqrt{(z-a)(z-b)(z-c)(z-d)}$ is a torus
  with four branch points.}

\section{Conclusion}

For any given 2D metric we have constructed kink spacetimes (inextendible if
$k\ne2$). In the presence of a Killing symmetry (thus covering e.g.\ all
generalized dilaton gravity solutions), furthermore, there occurred actually a
continuous one-parameter
family of solutions for each kink-number,
which were inextendible even for $k=2$. A geometrical interpretation of this
parameter has been provided in terms of holonomy resp.\ parallel displacement.
Although a complete classification of the maximal extensions of these kinks
proves to be elusive (due to ambiguities in the extension process), the
characteristic parameters have been pointed out. For cylindrical regions,
furthermore, a conformal but non-analytic as well as an analytic coordinate
system is given. As a by-product we have also found generalizations
of the bare kink and the Misner torus.

\section*{Acknowledgement}

The work has been supported in part by the Austrian Fonds zur F\"orderung
der wissenschaftlichen Forschung (FWF), project P10221-PHY.
T.S.\ is grateful also to the Erwin Schr\"odinger Institute, Vienna, for
an invitation which lead to fruitful discussions also on the topic of the
present paper.

\vspace{1cm}

\begin{appendix}
\renewcommand{\theequation}{A.\arabic{equation}}
\setcounter{equation}{0}
\section*{Appendix A: Conformal charts for flat kinks}

In this Appendix we provide conformal coordinate charts for the kink
manifolds of Fig.\ 6({\it a\/}),({\it b\/}).
\begin{figure}[htb]
 \begin{center}
 \leavevmode
 \epsfxsize 13cm \epsfbox{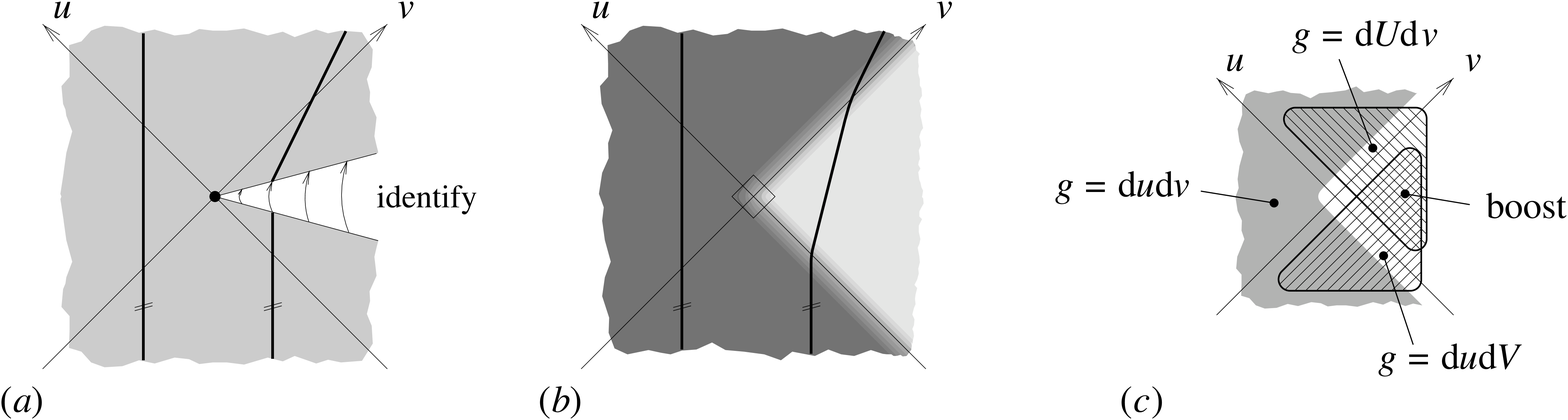}
 \end{center}
 \renewcommand{\baselinestretch}{.9}
 \small \normalsize
 \begin{quote}
 {\bf Figure 10:} {\small Minkowski kink with non-trivial holonomy.
   ({\it a\/}) shows once more the manifold Fig.\ 6({\it a\/}).
   Instead of this cutting procedure it may as well be described by an
   explicit (non-analytic) conformal chart \re{boostkink}), provided the
   puncture is enlarged to a small square. This is illustrated in ({\it b\/}),
   where the shading indicates the conformal factor (dark corresponding to a
   larger factor). ({\it c\/}) depicts some coordinate systems used in the
   text.}
 \end{quote}
\end{figure}
Let us start with the kink Fig.\ 6({\it a\/}) (cf.\ Fig.\
10). As already proved in section 3, it is impossible to give a smooth
conformal chart for the entire manifold. This problem can be circumvented,
however, if we enlarge the puncture to a square: Let $f(x)$ be a smooth
function which vanishes for $x<-1$ and equals 1 for $x>+1$. Then the metric
\begin{equation}
  g=e^{-\o f(-u)f(v)}{\rm d}u{\rm d}v
  \plabel{boostkink}
\end{equation}
is flat outside the square $-1<u,v<1$ (cf.\ Fig.\ 6({\it b\/}),
where the shading indicates the conformal factor). Still, the
holonomy is non-trivial, as is easily calculated, and it depends on the
parameter $\o$ in the exponent:
\footnote{However, even without calculation it is obvious that the conformal
  factor mimics the cutting procedure by giving less metric ``weight'' to the
  right-hand sector.}
Let
\begin{equation}
  F(x):=\int^x e^{-\o f(z)}{\rm d}z\,.
\end{equation}
Clearly $F(x)=x$ for $x<-1$, $F(x)=xe^{-\o}+const$ for $x>1$,
and for simplicity we assume that $f$ is fine-tuned such as to make
$const$ vanish. Note that the metric \re{boostkink}) is already in Minkowski
form $g={\rm d}u{\rm d}v$ throughout the dark shaded part of
Fig.\ 10({\it b\/}),({\it c\/}). There are two possibilities to transform
\re{boostkink}) into Minkowski form also on the right-hand sector: Introducing
$U(u):=-F(-u)$ instead of $u$ does this job for the upper right half, $v>1$,
and likewise replacing $v$ by $V(v):=F(v)$ works for the lower right half,
$u<-1$ (cf.\ Fig.\ 10({\it c\/})). However, on the right-hand sector itself
the two overlapping coordinate systems $(U,v)$ and $(u,V)$ disagree by a boost,
\begin{equation}
  {U\choose v}=
     \left(\begin{array}{cc}e^{-\o}&0\\0&e^\o\end{array}\right)
                                                      {u\choose V} \,,
\end{equation}
which proves the assertion.

In order to get the square cut out as small as possible (preferably
pointlike), the interval where $f$ ascends must be made
narrower. Thus, in the limiting case, one could describe the
entire solution --- allowing for a distributional conformal factor --- as
\begin{equation}
  g=e^{-\o \theta(-u)\theta(v)}{\rm d}u{\rm d}v\,,
\end{equation}
where $\theta(x)$ denotes the Heaviside step function.
The conformal factor then takes one constant value in the right-hand
(resp.\ any other) sector and another value everywhere else.
A short exposition of distributional metrics and their usefulness may be
found in \pcite{BalDistr}.

\begin{figure}[htb]
 \begin{center}
 \leavevmode
 \epsfxsize 13cm \epsfbox{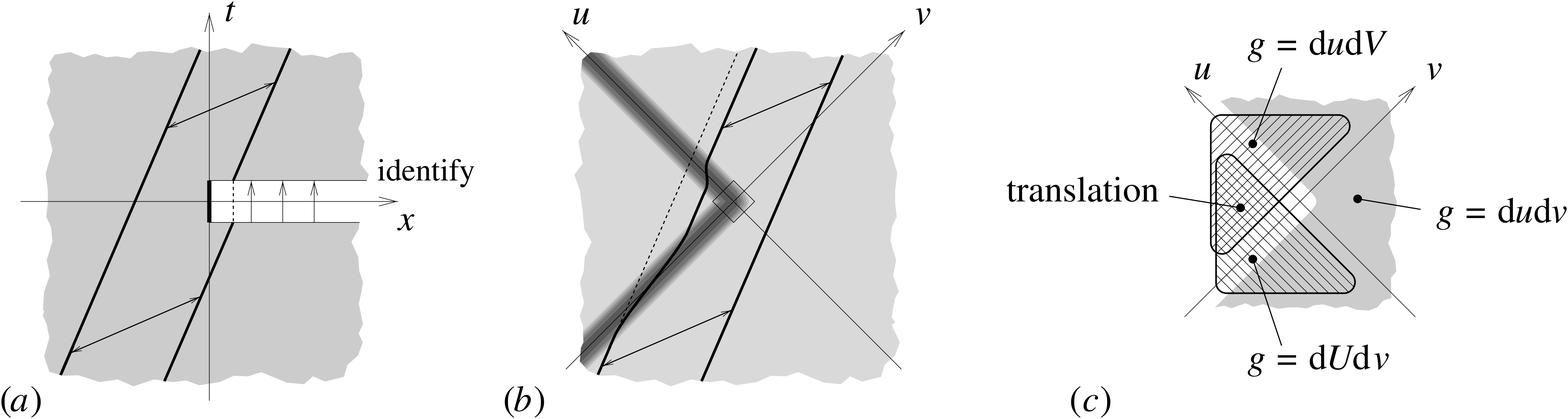}
 \end{center}
 \renewcommand{\baselinestretch}{.9}
 \small \normalsize
 \begin{quote}
 {\bf Figure 11:} {\small A similar construction can be applied to
   the ``translational'' Minkowski kink of Fig.\ 6({\it b\/}),
   leading to \re{noboostkink}).}
 \end{quote}
\end{figure}
For the Minkowski kink with trivial holonomy (Fig.\ 6({\it b\/}) or
Fig.\ 11({\it a\/})) an analogous chart can be found. The above reasoning
would suggest to imitate the cutting by suitably ``diluting'' the metric on
the strip in question. However, since the metric should remain
non-degenerate it seems wiser to turn the tables and {\em insert\/}
a strip of the same width $\o$ on the opposite side ($x<0$ in this example).
This can be achieved by putting
\begin{equation}
  g=\left(1+\o f(-x)f'(t)\right)^2{\rm d}t^2-{\rm d}x^2\,,
 \plabel{noboostkink1}
\end{equation}
with $f$ as before: Whereas on the right-hand side, $x>1$, \re{noboostkink1})
is in Minkowski form, $g={\rm d}t^2-{\rm d}x^2$, on the left-hand side,
$x<-1$, this form can only be attained when substituting $t$ by
$T(t):=t+\o f(t)$. But then, crossing the strip $-1<t<1$ clearly increases the
coordinate $T$ by $\o$ against the right-hand coordinate $t$, as desired.
A conformal chart is obtained when splitting this vertical shift into two
similar shifts of the null coordinates $u,v=t\mp x$
(the horizontal components cancelling, but the vertical ones adding up):
\begin{equation}
  g=\left(1+\o f(u)f'(v)+\o f(-v)f'(u)\right){\rm d}u{\rm d}v\,
 \plabel{noboostkink}
\end{equation}
This is illustrated in Fig.\ 11({\it b\/}). Again, introducing
$U(u):=u+\o f(u)$ on the lower left half, $v<-1$, and $V(v):=v-\o f(-v)$ on
the upper left half, $u>1$, allows us to extend the Minkowski metric from the
right-hand side into the left-hand sector, cf.\ Fig.\ 11({\it c\/}).
On the overlap the two coordinate systems are related by
\begin{equation}
  {U\choose v}= {u+\o\choose V+\o} \,,
\end{equation}
i.e.\ by a shift of length $\o$ into $t$-direction. [If the two terms in
\re{noboostkink}) were given different (positive) weights $\o_1$, $\o_2$,
with $\o_1 \o_2 = \o^2$, then the corresponding translation direction would
be boosted; sufficiently far from
the origin, however, the resulting kink-manifolds are still isometric.]
Also here the limit of the shrinking interval can be taken, but this time
it will involve a $\delta$-distribution,
$f'(x) \rightarrow \theta'(x)=\delta(x)$. While this does not pose any
problems for \re{noboostkink}), the metric \re{noboostkink1})
looks rather undefined then, as it contains a term $\delta(t)^2$.

\section*{Appendix B: Extension of the ``translation''-kink}

Here we want to show how the flat kink of Fig.\ 6({\it b\/})
resp.\ Fig.\ 11({\it a\/}) can be extended further.
A similar procedure may be applied to all kinks obtained in Sec.\ 4.
In these examples a strip has been removed, leaving an incomplete edge
(bold line in Fig.\ 12({\it a\/})).
\begin{figure}[htb]
 \begin{center}
 \leavevmode
 \epsfxsize 14cm \epsfbox{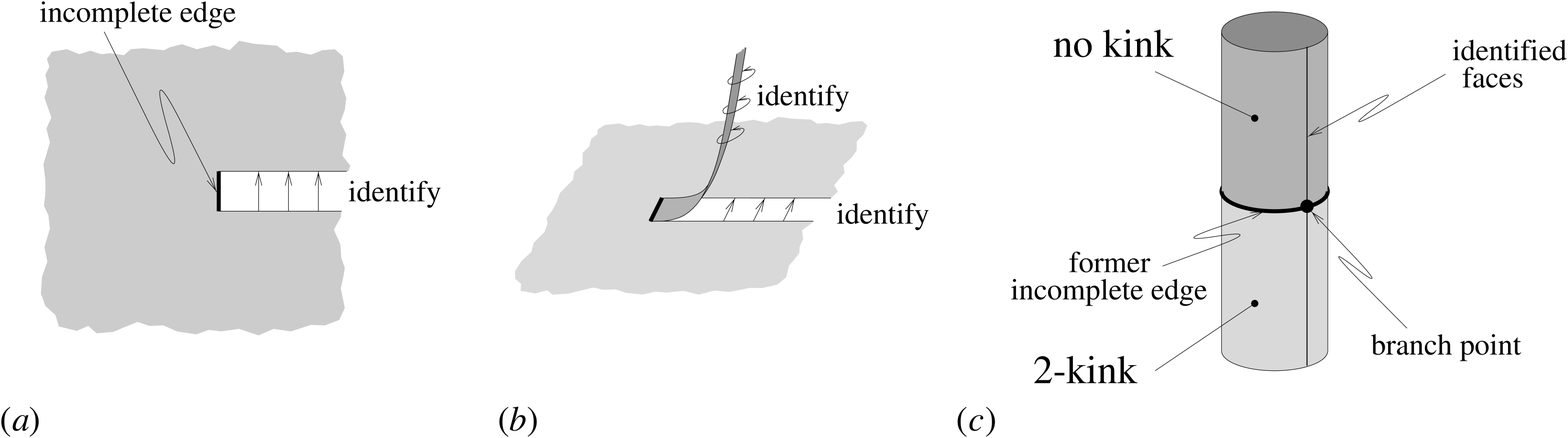}
 \end{center}
 \renewcommand{\baselinestretch}{.9}
 \small \normalsize
 \begin{quote}
 {\bf Figure 12:} {\small Possible maximal extension of the
    ``translation''-kinks.}
 \end{quote}
\end{figure}
However, one can simply extend the manifold beyond the edge into a new
layer: The perhaps easiest way to look at this is to keep the previously
removed strip attached to the rest ({\it b\/}) and to sew
together its faces, too. The result is then one cylinder ({\it c\/}), half of
which is the original 2-kink solution, the other half having no kink.
(In this example there are closed timelike curves in the latter half
cylinder; they would not occur, if we had chosen a spacelike translation.)
The identified endpoints of the previously incomplete edge constitute a
branch point which should be
removed. Certainly, to obtain the universal covering one has now first to
unwrap this cylinder (thus introducing infinitely many copies of the
branch point) and subsequently also to unwrap the manifold around those
branch points into new overlapping layers (thereby once again multiplying
the number of branch points).

\end{appendix}

\end{document}